\documentclass[12pt]{article}

\usepackage[letterpaper,top=2cm,bottom=2cm,left=2cm,right=2cm,marginparwidth=1.75cm]{geometry}
\usepackage{amsmath,amssymb}
\usepackage{graphicx}
\usepackage[colorlinks=true, allcolors=blue]{hyperref}
\usepackage{authblk}
\usepackage{tabularray}
\usepackage{bbm}
\usepackage{multirow}
\usepackage{cite}
\UseTblrLibrary{booktabs}

\def\mathswitchr#1{\relax\ifmmode{\mathrm{#1}}\else$\mathrm{#1}$\fi}
\def\mathswitch#1{\relax\ifmmode#1\else$#1$\fi}
\newcommand{\brc}[1]{\left(#1\right)}

\newcommand{\corr}{\mathswitchr{Corr}}
\newcommand{\bbeta}{\mathswitch{\boldsymbol{\beta}}}
\newcommand{\bgam}{\mathswitch{\boldsymbol{\gamma}}}

\newcommand{\bC}{\mathswitch{\boldsymbol{C}}}
\newcommand{\bI}{\mathswitch{\boldsymbol{\mathrm{I}}}}

\newcommand{\bU}{\mathswitch{\boldsymbol{U}}}

\newcommand{\bZ}{\mathswitch{\boldsymbol{Z}}}

\newcommand{\declare}[2]{\vspace{2em}\noindent{\fontsize{14}{14}\selectfont\textbf{#1}}{%
\par\vspace{3pt}{\fontsize{12}{14}\selectfont #2}\par}}

\providecommand{\keywords}[1]
{
  \small
  \noindent
  \textbf{Keywords:} #1
}

\title{\textbf{\Large Testing Risk Difference of Two Proportions for Combined Unilateral and Bilateral Data}}
\author{Jia Zhou \footnote{\href{mailto:jiazhou@buffalo.edu}{jiazhou@buffalo.edu}} }
\author{Chang-Xing Ma \footnote{\href{mailto:cxma@buffalo.edu}{cxma@buffalo.edu}}}
\affil{Department of Biostatistics, University at Buffalo, Buffalo, NY 14214, USA}
\date{}

\begin{document}
\maketitle

\begin{abstract}
\fontsize{12pt}{14pt}\selectfont
In clinical studies with paired organs, binary outcomes often exhibit intra-subject correlation and may include a mixture of unilateral and bilateral observations. Under Donner’s constant correlation model, we develop three likelihood-based test statistics (the likelihood ratio, Wald-type, and score tests) for assessing the risk difference between two proportions. Simulation studies demonstrate good control of type I error and comparable power among the three tests, with the score test showing slightly better stability. Applications to otolaryngologic and ophthalmologic data illustrate the methods. An online calculator is also provided for power analysis and risk difference testing. The score test is recommended for practical use and future studies with combined unilateral and bilateral binary data.

\vspace{5mm}
\end{abstract}

\section{Introduction}
\label{sec:intro}
In comparative clinical studies, binary outcomes are commonly used to assess treatment efficacy or disease occurrence, where each observation represents a dichotomous event such as success/failure or cured/uncured. In many medical applications, particularly those involving paired organs, such as eyes and ears, responses are measured bilaterally. In practice, however, datasets often contain a mixture of unilateral and bilateral observations, as some subjects contribute paired organ data while others provide only a single measurement. Such imbalance may arise due to prior surgical removal, congenital absence, or missing measurements in one organ. Analyzing these combined unilateral and bilateral data jointly, rather than excluding unilateral cases, is essential to avoid loss of information and to reduce potential bias in statistical inference. 

A key challenge in analyzing bilateral outcomes lies in appropriately accounting for the intra-subject correlation between paired organs. Several statistical models have been proposed to handle this correlation. Rosner~\cite{Rosner_1982} introduced the ``constant $R$'' model, which assumes equal dependence between paired organs within an individual and tests for equality of proportions across treatment groups. Dallal~\cite{Dallal_1988} identified limitations of Rosner's model when the marginal prevalences differ substantially between groups and proposed an alternative model assuming a constant conditional probability of response in one organ given the other. Donner~\cite{Donner1989rhoModel} later developed a constant correlation model that parameterizes the intra-subject dependence through a single correlation coefficient $\rho$, which Thompson~\cite{Thompson_1993} demonstrated to be robust through simulation studies. Moreover, Clayton~\cite{clayton1978model} proposed a general model for association in bivariate data using the Clayton copula, which expresses the joint distribution of two variables in terms of their marginal cumulative distribution functions and a dependence parameter. Although originally developed for bivariate survival or event-time data, the Clayton copula model can be extended to correlated binary outcomes and has been applied in paired organ studies to capture intra-subject dependence~\cite{liang2025testing,zhou2025goodness}.

Building on these correlation structures, numerous likelihood-based methods have been developed for inference with correlated binary data. For instance, Tang~\textit{et al.}~\cite{Tang_2008} and Pei~\textit{et al.}~\cite{Pei_2008} proposed asymptotic methods for testing equality of proportions between two groups. Ma and others~\cite{ma2017rho,ma2022testing} investigated three asymptotic tests (likelihood ratio, Wald-type and score) for testing homogeneity of proportions among $g\ge2$ groups. Other studies have considered relative measures, such as relative risk and odds ratio, and their interval estimations have been studied under different correlation models, with or without stratification (e.g., see~\cite{li2022statistical,tian2025confidence}).

While tests for equality of proportions have been extensively studied, practical research questions often concern the difference in risk (or proportion) between two treatment groups, referred to as the \textit{risk difference}. Compared with relative measures such as odds ratio or relative risk, the risk difference provides a more direct and interpretable measure of treatment effect, especially in clinical and public health contexts where absolute risk reduction is of primary interest~\cite{noordzij2017relative}. It also forms the basis for important inferential quantities such as the number needed to treat (NNT)~\cite{cook1995number,mcalister2008number}. Several studies have examined inference procedures for the risk difference in correlated binary data. Pei~\textit{et al.}~\cite{pei2012confidence} compared three methods (profile likelihood, asymptotic score and Wald-type with three estimated variances) for constructing confidence intervals for differences in proportions in a two-arm randomized trial. More recently, Sun~\textit{et al.}~\cite{sun2025interval} studied various confidence intervals of common risk difference in stratified unilateral and bilateral data based on the Dallal’s model. In this paper, we focus on testing the risk difference between two proportions under Donner’s constant $\rho$ model. We derive three likelihood-based test statistics: the likelihood ratio, Wald-type, and score tests, respectively, to evaluate their performance through extensive simulation studies. To facilitate practical implementation, we also develop an online calculator for power analysis and hyptohesis testing using user-supplied data.

The rest of the paper is organized as follows. Section~\ref{sec:methods} presents the construction of the three likelihood-based test statistics under Donner's model. Section~\ref{sec:results} reports numerical results, including simulation studies (Section~\ref{sec:results:simulation}) evaluating empirical type I error rates and powers, and two real-world applications in otolaryngologic and ophthalmologic studies (Section~\ref{sec:results:real}). Section~\ref{sec:conclusions} concludes the paper.

\section{Methods}
\label{sec:methods}
Let $Z_{ijk}$ denote the response of the $k$-th ($k=1,2$) paired organ of the $j$-th subjet in the $i$-th ($i=1,2$) group, where $Z_{ijk}=1$ indicates the occurence of a particular condition (e.g., the organ is cured or affected) and $Z_{ijk}=0$ otherwise. For the unilateral portion of the date, each $Z_{ijk}$ ($k=1$ or $k=2$) follows a Bernoulli distribution. For the bilateral portion of the data, the vector $\bZ_{ij}=\brc{Z_{ij1},Z_{ij2}}^T$ follows a paired Bernoulli distribution. Let $m_{ri}$ denote the number of bilateral subjects in the $i$-th group who have $r~\brc{r=0,1,2}$ organs cured (or affected), and let $n_{ri}$ denote the number of unilateral subjects in the $i$-th group who have $r~\brc{r=0,1}$ organs cured (or affected). Then,  
\begin{align*}
  &m_{ri}=\sum_{j=1}^{m_{+i}}I\brc{Z_{ij1}+Z_{ij2}=r}, \quad r=0,1,2, \\
  &n_{ri}=\sum_{j=1}^{n_{+i}}I\brc{Z_{ijk}=r}, \quad r=0,1;~k=1\text{ or }2,
\end{align*}
where $m_{+i}=m_{0i}+m_{1i}+m_{2i}$ and $n_{+i}=n_{0i}+n_{1i}$ are the numbers of the subjects contributing bilateral and unilateral observations in the $i$-th group, respectively. The data structure can be summarized as a combination of a $3\times2$ contingency table for the bilateral portion and a $2\times2$ contingency table for the unilateral portion, as shown in Table~\ref{tab:data_struc}.  
\begin{table}[thpb]
    \centering
    \caption{Frequency table for number of cured (or affected) organs for subjects in $g$ groups.}
    \label{tab:data_struc}
    \begin{tabular}{cccc}
    \toprule
    \# of cured or affected organs &group 1 &group 2 &total \\
    \midrule
    0 &$m_{01}$ &$m_{02}$ &$m_{0+}$ \\
    1 &$m_{11}$ &$m_{12}$ &$m_{1+}$ \\
    2 &$m_{21}$ &$m_{22}$ &$m_{2+}$ \\
    total &$m_{+1}$ &$m_{+2}$ &$m_{++}$ \\
    \midrule
    0 &$n_{01}$ &$n_{02}$ &$n_{0+}$ \\
    1 &$n_{11}$ &$n_{12}$ &$n_{1+}$ \\
    total &$n_{+1}$ &$n_{+2}$ &$n_{++}$ \\
    \bottomrule
    \end{tabular}
\end{table}

The proportion of organs that are cured (or affected) in the $i$-th group is assumed to be $Pr\brc{Z_{ijk}=1}=\pi_i$. Under Donner's constant $\rho$ model~\cite{Donner1989rhoModel}, the intra-subject correlation satisfies $\corr\brc{Z_{ijk},Z_{ij,3-k}}=\rho$. Therefore, given $m_{+i}$ and $n_{+i}$ in the $i$-th group, the counts $\brc{m_{0i},m_{1i},m_{2i}}$ follow a trinomial distribution, and $\brc{n_{0i},n_{1i}}$ follow a binomial distribution, 
\begin{equation}
  \brc{m_{0i},m_{1i},m_{2i}}\sim Trinomial\brc{m_{+i},p_{0i},p_{1i},p_{2i}}, \quad
  n_{1i} \sim Binomial\brc{n_{+i},\pi_i}, 
  \label{eq:distribution}
\end{equation}
where the joint probabilities $p_{ri}$'s ($r=0,1,2$) are given by
\begin{equation}
    \begin{aligned}
        &p_{2i}=Pr\brc{Z_{ij1}=1,Z_{ij2}=1}=E\brc{Z_{ij1}Z_{ij2}}=\pi_i\brc{\pi_i+\brc{1-\pi_i}\rho}, \\
        &p_{1i}=\sum_{k=1}^2Pr\brc{Z_{ijk}=1,Z_{ij,3-k}=0}=2\brc{\pi_i-p_{2i}}=2\pi_i\brc{1-\pi_i}\brc{1-\rho}, \\
        &p_{0i}=Pr\brc{Z_{ij1}=0,Z_{ij2}=0}=1-p_{1i}-p_{2i}=\brc{1-\pi_i}\brc{1-\pi_i+\pi_i~\rho}.  
    \end{aligned}
    \label{eq:prob_ri}
\end{equation}

Consequently, the log-likelihood as a function of $\bbeta=\brc{\pi_1,\pi_2,\rho}^T$ can be written as
\begin{equation}
  \begin{aligned}
    \ell\brc{\bbeta}=&\sum_{i=1}^2\bigg\{m_{0i}\log\left[\brc{1-\pi_i}\brc{1-\pi_i+\pi_i\rho}\right]
    + m_{1i}\log\left[2\pi_i\brc{1-\pi_i}\brc{1-\rho}\right] \\
    & + m_{2i}\log\left[\pi_i\brc{\pi_i+\brc{1-\pi_i}\rho}\right] 
    + n_{0i}\log\brc{1-\pi_i}+n_{1i}\log\brc{\pi_i}\bigg\} +\text{const}, 
  \end{aligned}
  \label{eq:loglik:1}
\end{equation}
where the term `const" denotes a constant that depends only on the counts $m_{ri}$ and $n_{ri}$.

Likewise, by reparameterizing in terms of the risk difference $\delta=\pi_2-\pi_1$, the log-likelihood as a function of $\bgam=\brc{\delta,\pi_1,\rho}^T$ can be expressed as
\begin{equation}
  \begin{aligned}
    \ell_1\brc{\bgam}=~&m_{01}\log\left[\brc{1-\pi_1}\brc{1-\pi_1\brc{1-\rho}}\right]
    + m_{02}\log\left[\brc{1-\delta-\pi_1}\brc{1-\brc{\delta+\pi_1}\brc{1-\rho}}\right] \\
    & + m_{11}\log\left[2\pi_1\brc{1-\pi_1}\brc{1-\rho}\right]
    + m_{12}\log\left[2\brc{\delta+\pi_1}\brc{1-\delta-\pi_1}\brc{1-\rho}\right] \\
    & + m_{21}\log\left[\pi_1\brc{\pi_1+\brc{1-\pi_1}\rho}\right]
    + m_{22}\log\left[\brc{\delta+\pi_1}\brc{\delta+\pi_1+\brc{1-\delta-\pi_1}\rho}\right] \\
    & + n_{01}\log\brc{1-\pi_1} + n_{02}\log\brc{1-\delta-\pi_1}
    + n_{11}\log\brc{\pi_1} + n_{12}\log\brc{\delta+\pi_1}
    + \text{const}. 
  \end{aligned}
  \label{eq:loglik:2}
\end{equation}

Our primary interest is to test whether the risk difference between the two groups equals a pre-specified constant. Formally, the hypotheses are 
\begin{equation}
  H_0:~\delta=\delta_0
  \quad
  \text{vs}
  \quad
  H_1:~\delta\ne\delta_0,
  \label{eq:hypotheses}
\end{equation}
for some given constant $\delta_0$.

In what follows, we derive the unconstrained maximum likelihood estimates (MLEs) of the parameters $\bgam=\brc{\delta,\pi_1,\rho}^T$, as well as the constrained MLEs of $\brc{\pi_1,\rho}$ under the null hypothesis $H_0:\delta=\delta_0$. Based on these estimates, we then construct three likelihood-based test statistics (likelihood ratio, Wald-type and score) to assess the hypothesis.

\subsection{Maximum Likelihood Estimates}
\label{sec:methods:mle}
\subsubsection{Unconstrained MLEs}
The unconstrained MLEs of $\brc{\delta,\pi_1,\rho}$ can be derived from the global MLEs of $\bbeta=\brc{\pi_1,\pi_2,\rho}^T$ by using $\hat{\delta}=\hat{\pi}_2-\hat{\pi}_1$. 
The global MLEs of $\bbeta=\brc{\pi_1,\pi_2,\rho}^T$ can be obtained by solving the following normal equations
\begin{subequations}
  \begin{equation}
    \begin{aligned}
      \frac{\partial\ell}{\partial\pi_i}=~&\frac{m_{0+}\brc{2\brc{1-\rho}\pi_i+\rho-2}}{\brc{1-\pi_i}\brc{1-\brc{1-\rho}\pi_i}}+\frac{m_{1+}\brc{1-2\pi_i}}{\pi_i\brc{1-\pi_i}}+\frac{m_{2+}\brc{2\brc{1-\rho}\pi_i+\rho}}{\pi_i\brc{\rho\brc{1-\pi_i}+\pi_i}} \\
      &-\frac{n_{0+}}{1-\pi_i}+\frac{n_{1+}}{\pi_i}=0,
      \quad
      \brc{i=1,2}, 
    \end{aligned}
    \label{eq:norm:mle1:donner:1}
  \end{equation}
  \begin{equation}
    \frac{\partial\ell}{\partial\rho}=\sum_{i=1}^2\left[\frac{m_{0i}\pi_i}{1-\brc{1-\rho}\pi_i}+\frac{m_{2i}\brc{1-\pi_i}}{\rho\brc{1-\pi_i}+\pi_i}\right]-\frac{m_{1+}}{1-\rho}=0. 
    \label{eq:norm:mle1:donner:2}
  \end{equation}
\end{subequations}

The equation (\ref{eq:norm:mle1:donner:1}) leads to a cubic equation
\begin{equation}
    a_i\pi_i^3+b_i\pi_i^2+c_i\pi_i+d_i=0, 
    \label{eq:cubic:pi:donner}
\end{equation}
for $i=1,2$, where 
\begin{align*}
    &a_i=\brc{1-\rho}^2\brc{2m_{+i}+n_{+i}}, \\
    &b_i=\brc{1-\rho}\brc{\brc{3\rho-2}m_{0i}+3\brc{\rho-1}m_{1i}+\brc{3\rho-4}m_{2i}+\brc{\rho-1}n_{0i}+2\brc{\rho-1}n_{1i}}, \\
    &c_i=\rho\brc{\rho-2}m_{0i}+\brc{\brc{\rho\brc{\rho-4}+1}m_{1i}+\brc{\rho\brc{\rho-4}+2}m_{2i}+\brc{\rho\brc{\rho-3}+1}n_{1i}}, \\
    &d_i=\rho\brc{m_{1i}+m_{2i}+n_{1i}}. 
\end{align*}

It can be shown that the analytic root for the cubic equation (\ref{eq:cubic:pi:donner})~\cite{ma2022testing},
\begin{equation}
  \hat{\pi}_i=\frac{-b_i+2\sqrt{\Delta_0}\cos\brc{\frac{\theta}{3}-\frac{2\pi}{3}}}{3a_i},
  \label{eq:cubic:rt}
\end{equation}
yields the maximum of the log-likelihood in (\ref{eq:loglik:1}) for a given $\rho$, where
$$
\theta=\cos^{-1}\brc{-\frac{\Delta_1}{2\Delta_0^{3/2}}},
$$
and
$$
\Delta_0=b_i^2-3a_ic_i, \quad
\Delta_1=2b_i^3-9a_ib_ic_i+27a_i^2d_i.
$$
The global MLEs of $\pi_i$ and $\rho$ can be obtained through the iterative procedure outlined as follows.
\begin{enumerate}
\item[1.] \textbf{Update} $\pi_i$: at the $t$-th step, for a given $\hat{\rho}^{(t)}$, compute $\hat{\pi}_i^{(t)}$ using equation (\ref{eq:cubic:rt}).
\item[2.] \textbf{Update} $\rho$: at the $\brc{t+1}$-th step, update $\hat{\rho}^{(t+1)}$ via the Newton-Raphson method:
  $$
  \hat{\rho}^{(t+1)}=\hat{\rho}^{(t)}+\brc{\left.-\frac{\partial^2\ell\brc{\bbeta}}{\partial\rho^2}\right|_{\bbeta=\hat{\bbeta}^{(t)}}}^{-1}\left.\frac{\partial\ell\brc{\bbeta}}{\partial\rho}\right|_{\bbeta=\hat{\bbeta}^{(t)}},
  $$
  where $\partial\ell\brc{\bbeta}/\partial\rho$ corresponds to the middle expression in equation (\ref{eq:norm:mle1:donner:2}), and $\partial^2\ell\brc{\bbeta}/\partial\rho^2$ is the associated Hessian term and takes the following form
  $$
  \frac{\partial^2\ell\brc{\bbeta}}{\partial\rho^2}=\sum_{i=1}^2\left[-\frac{m_{0i}\pi_i^2}{\brc{1-\brc{1-\rho}\pi_i}^2}+\frac{m_{1i}}{\brc{1-\rho}^2}+\frac{m_{2i}\brc{1-\pi_i}^2}{\brc{\rho\brc{1-\pi_i}+\pi_i}^2}\right].
  $$
\item[3.] \textbf{Check convergence}: repeat \textbf{Steps 1 - 2} until $\hat{\rho}$ converges, as measured by
  $$\delta^{(t+1)}=\left|\hat{\rho}^{(t+1)}-\hat{\rho}^{(t)}\right|.$$
  The iteration stops when $\delta^{(t+1)}<\delta_c$ for a sufficiently small tolerance, such as $\delta_c=10^{-6}$. 
\end{enumerate}

After obtaining $\hat{\bbeta}$, the MLE of the risk difference $\delta$ is $\hat{\delta}=\hat{\pi}_2-\hat{\pi}_1$.

\subsubsection{Constrained MLEs}
The constrained MLEs of $\pi_1$ and $\rho$ under $H_0:~\delta=\delta_0$ can be obtained by solving the following normal equations, 
\begin{subequations}
  \begin{equation}
    \begin{aligned}
    \left.\frac{\partial\ell_1}{\partial\pi_1}\right|_{\delta=\delta_0}\!\!\!\!\!\!\!\!=~&
    -\frac{m_{01}+m_{11}+n_{01}}{1-\pi_1}+\frac{m_{11}+m_{21}+n_{11}}{\pi_1}
    -\frac{m_{02}+m_{12}+n_{02}}{1-\delta_0-\pi_1}
    +\frac{m_{12}+m_{22}+n_{12}}{\delta_0+\pi_1} \\
    &-\frac{m_{01}\brc{1-\rho}}{1-\brc{1-\rho}\pi_1}
    +\frac{m_{21}\brc{1-\rho}}{\rho+\brc{1-\rho}\pi_1}
    -\frac{m_{02}\brc{1-\rho}}{1-\brc{1-\rho}\brc{\delta_0+\pi_1}}
    +\frac{m_{22}\brc{1-\rho}}{\rho+\brc{1-\rho}\brc{\delta_0+\pi_1}} \\
    =&0, 
    \end{aligned}
    \label{subeq:norm:1}
  \end{equation}
  \begin{equation}
    \begin{aligned}
      \left.\frac{\partial\ell_1}{\partial\rho}\right|_{\delta=\delta_0}\!\!\!\!\!\!\!\!=~&
      \frac{1}{1-\rho}\brc{
        \frac{m_{01}}{1-\brc{1-\rho}\pi_1}
        +\frac{m_{21}}{\rho+\brc{1-\rho}\pi_1}
        +\frac{m_{02}}{1-\brc{1-\rho}\brc{\delta_0+\pi_1}}
        +\frac{m_{22}}{\rho+\brc{1-\rho}\brc{\delta_0+\pi_1}}
      } \\
      &-\frac{m_{++}}{1-\rho} \\
      =&0.
    \end{aligned}
    \label{subeq:norm:2}
  \end{equation}
  \label{eq:norm:h0}
\end{subequations}

Since there is no closed-form solution for $\pi_1$ or $\rho$ from equations (\ref{subeq:norm:1} -- \ref{subeq:norm:2}), we compute the constrained MLEs using the Fisher scoring method. Specifically, given the estimates $\brc{\hat{\pi}_1^{(t)},\hat{\rho}^{(t)}}^T$ at the $t$-th step, the updates at the $(t+1)$-th step are obtained as 
$$
\brc{\hat{\pi}_1^{(t+1)},\hat{\rho}^{(t+1)}}^T=
\brc{\hat{\pi}_1^{(t)},\hat{\rho}^{(t)}}^T
+\bI_0\brc{\hat{\pi}_1^{(t)},\hat{\rho}^{(t)}}^{-1}\bU\brc{\hat{\pi}_1^{(t)},\hat{\rho}^{(t)}}
$$
where $\bU\brc{\pi_1,\rho}=\brc{\partial\ell_1/\partial\pi_1,\partial\ell_1/\partial\rho}_{\delta=\delta_0}^T$ is the score function, and $\bI_0\brc{\pi_1,\rho}^{-1}$ is the inverse of the $2\times2$ Fisher information matrix $\bI_0\brc{\pi_1,\rho}$ with respect to $\pi_1$ and $\rho$, which coincides with the lower-right block of the $3\times3$ Fisher information matrix $\bI\brc{\delta=\delta_0,\pi_1,\rho}$, i.e., 
\begin{equation}
  \bI\brc{\pi_1,\rho}=
  \brc{
  \begin{array}{cc}
    I_{22} &I_{23} \\
    I_{32} &I_{33}
  \end{array}
  }_{\delta=\delta_0},
\end{equation}
with $I_{22},I_{23},I_{32},I_{33}$ the elements in $\bI\brc{\delta,\pi_1,\rho}$ provided in Appendix~\ref{app:FI}. 

This updating process is repeated until convergence, which is assessed by the change in successive estimates falling below a pre-specified tolerance. Specifically, let
$$
\delta^{(t+1)}=\sqrt{\brc{\hat{\pi}_1^{(t+1)}-\hat{\pi}_1^{(t)}}^2+\brc{\hat{\rho}^{(t+1)}-\hat{\rho}^{(t)}}^2}, 
$$
and stop updating when $\delta^{(t+1)}<\delta_c$ for a sufficient small tolerance, such as $\delta_c=10^{-6}$.

\subsection{Likelihood-based Test Statistics}
\label{sec:methods:tests}
Based on the unconstrained and constrained MLEs, we construct three likelihood-based test statistics: (i) the likelihood ratio, (ii) Wald-type, and (iii) score tests, for testing the hypotheses in (\ref{eq:hypotheses}).
Let $\hat{\bgam}=\brc{\hat{\delta},\hat{\pi}_1,\hat{\rho}}$ and $\hat{\bgam}_0=\brc{\hat{\pi}_{1,0},\hat{\rho}_0}$ be the unconstrained and constrained MLEs, respectively. Under the null hypothesis $H_0:~\delta=\delta_0$, the three test statistics are defined as follows.

\subsubsection*{(i) Likelihood Ratio Test}
The likelihood ratio (LR) test is given by 
\begin{equation}
  Q_{LR}=2\left[\ell_1\brc{\hat{\bgam}}-\ell_1\brc{\hat{\bgam}_0}\right].
  \label{eq:LR}
\end{equation}

According to Wilks' theorem, under $H_0$ the LR test asymptotically follows the chi-square distribution with one degree of freedom.

\subsubsection*{(ii) Wald-type Test}
The Wald-type statistic is constructed based on the asymptotic normality of the MLEs,
$$
\hat{\bgam}-\bgam\sim AN\brc{\boldsymbol{0},\bI\brc{\bgam}^{-1}}.
$$
Let $\bC^T=\brc{1,0,0}$, such that under $H_0:~\bC^T\bgam=\delta_0$. Then, the Wald-type test statistic is given by
\begin{equation}
  Q_W=\left.\brc{\bgam^T\bC-\delta_0}\left[\bC^T\bI\brc{\bgam}^{-1}\bC\right]^{-1}\brc{\bC^T\bgam-\delta_0}\right|_{\bgam=\hat{\bgam}}
  =\brc{\hat{\delta}-\delta_0}^2/I^{11}\brc{\hat{\bgam}}, 
  \label{eq:wald}
\end{equation}
where $I^{11}\brc{\hat{\bgam}}=\brc{\bI\brc{\hat{\bgam}}^{-1}}_{11}$ denotes the $\brc{1,1}$-th element of the inverse Fisher information matrix $\bI\brc{\hat{\bgam}}$. 
For a given $3\times3$ Fisher information matrix $\bI\brc{\bgam}=\brc{I_{ij}}_{i,j=1}^3$, it is straightforward to show that 
\begin{equation}
  I^{11}\brc{\bgam}=\brc{I_{11}
    -\brc{I_{12},I_{13}}
    \brc{
      \begin{array}{cc}
        I_{22} &I_{23} \\
        I_{32} &I_{33}
      \end{array}
    }^{-1}
    \brc{
      \begin{array}{c}
        I_{12} \\
        I_{13}
      \end{array}
    }
  }^{-1}.
  \label{eq:invI11}
\end{equation}
The explicit form of $\bI\brc{\bgam}$ is provided in Appendix~\ref{app:FI}.
The Wald-type test in (\ref{eq:wald}) asymptotically follows the chi-square distribution with one degree of freedom.

\subsubsection*{(iii) Score Test}
The score test under $H_0:~\delta=\delta_0$ is defined as 
\begin{equation}
  Q_S=\left.\bU\brc{\bgam}\bI\brc{\bgam}^{-1}\bU^T\brc{\bgam}\right|_{\bgam=\brc{\delta_0,\hat{\bgam}_0}},
  \label{eq:score}
\end{equation}
where $\bU\brc{\bgam}=\brc{\frac{\partial\ell_1}{\partial\delta},\frac{\partial\ell_1}{\partial\pi_1},\frac{\partial\ell_1}{\partial\rho}}$ is the score function, and $\bI\brc{\bgam}^{-1}$ is the inverse of the Fisher information matrix $\bI\brc{\bgam}$ given in Appendix~\ref{app:FI}. When evaluated at $\bgam=\brc{\delta_0,\hat{\bgam}_0}$, the score function reduces to $\bU\brc{\bgam=\brc{\delta_0,\hat{\bgam}}}=\brc{\left.\frac{\partial\ell_1}{\partial\delta}\right|_{\bgam=\brc{\delta_0,\hat{\bgam}_0}},0,0}$,
and the score test simplifies to
\begin{equation}
  Q_S=I^{11}\brc{\bgam}\left.\brc{\frac{\partial\ell_1}{\partial\delta}}^2\right|_{\bgam=\brc{\delta_0,\hat{\bgam}_0}},
  \label{eq:score:red}
\end{equation}
where $I^{11}\brc{\bgam}$ is defined in (\ref{eq:invI11}), and
$$
\frac{\partial\ell_1}{\partial\delta}=-\frac{m_{02}+m_{12}+n_{02}}{1-\delta-\pi_1}+\frac{m_{12}+m_{22}+n_{12}}{\delta+\pi_1}-\frac{m_{02}\brc{1-\rho}}{1-\brc{1-\rho}\brc{\delta+\pi_1}}+\frac{m_{22}\brc{1-\rho}}{\rho+\brc{1-\rho}\brc{\delta+\pi_1}}. 
$$
Likewise, the score test $Q_S$ asymptotically follows the chi-square distribution with one degree of freedom.

\section{Results}
\label{sec:results}

\subsection{Simulation Studies}
\label{sec:results:simulation}
We conduct simulation studies to evaluate the performance of the three likelihood-based test statistics proposed in Section~\ref{sec:methods:tests} by investigating their empirical type I error rates (TIEs) and powers. 
The following parameter settings are used in common for both empirical type I error and power evaluations:
$\rho=0,0.5,0.9$, $\pi_1=0.1,0.2,0.3$, $m_{+1}=m_{+2}=m=50,100,150$ and $n_{+1}=n_{+2}=n=50,100,150$.

\subsubsection{Empirical Type I Error}
\label{sec:results:simulation:TIE}
Under the null hypothesis $H_0:~\delta=\delta_0$ with $\delta_0=0.1,0.2,0.3$, and for given values of $\rho$, $\pi_1$, $m$, and $n$, datasets are randomly generated according to the distributions in (\ref{eq:distribution}). The three likelihood-based test statistics, $Q_{LR}$, $Q_W$, and $Q_S$, respectively defined in (\ref{eq:LR}), (\ref{eq:wald}) and (\ref{eq:score:red}), are then computed. The null hypothesis $H_0$ is rejected if $Q_i>\chi^2_{1-\alpha,1}$, where $i=LR,W,S$ indicates the test type, and $\chi^2_{1-\alpha,1}$ denotes the $\brc{1-\alpha}$-quantile of the chi-square distribution with one degree of freedom. Each simulation scenario is replicated $N=100,000$ times, and the empirical TIE is computed as the proportion of rejections among the total number of replications.

Tables~\ref{tab:TIE:1} - \ref{tab:TIE:3} present the empirical TIEs at the nominal significance level $\alpha=0.05$ for $\rho=0$, $0.5$, and $0.9$, respectively. Following the notation in Tang \textit{et al.}~\cite{Tang_2006}, a test statistic is considered \textit{robust} if $0.8\le\mathrm{TIE}/\alpha\le1.2$, \textit{liberal} if $\mathrm{TIE}/\alpha>1.2$, and \textit{conservative} if $\mathrm{TIE}/\alpha<0.8$. As can been seen, all the TIEs presented in the three tables for the three test statistics $Q_{LR}$, $Q_W$ and $Q_S$ fall within the \textit{robust} range $\left[0.04,0.06\right]$ when $\alpha=0.05$, indicating that all three tests maintain good control of the type I error across different parameter setting scenarios.

\begin{figure}[htpb]
  \centering
  \includegraphics[scale=0.6]{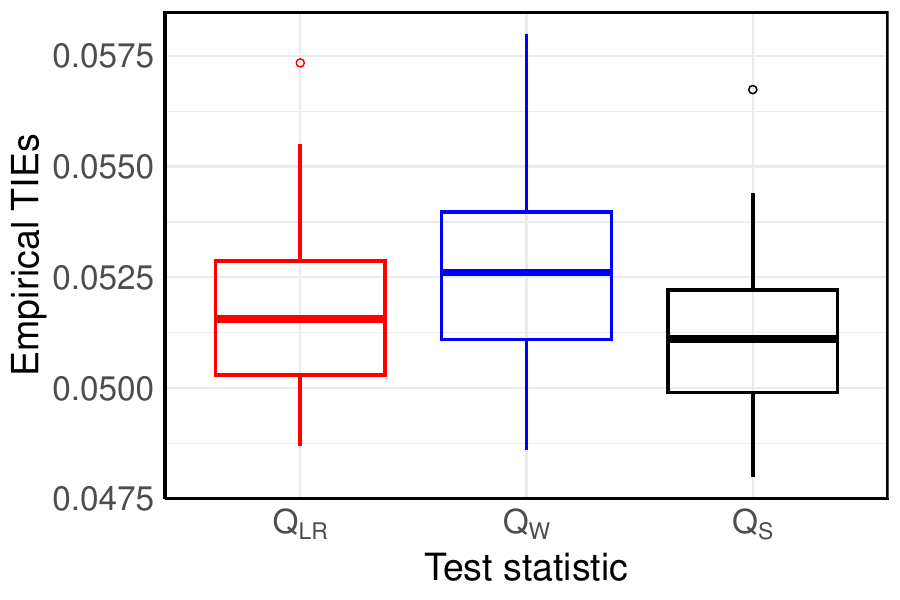}
  \caption{Boxplots of empirical TIEs for the likelihood ratio test ($Q_{LR}$), Wald-type test ($Q_W$) and score test ($Q_S$), respectively, based on $10,000$ randomly generated parameter configurations.}
  \label{fig:boxplot}
\end{figure}

To further assess the reliability of the proposed test statistics, we randomly generate $10,000$ parameter configurations within the following ranges: $\delta_0\in\brc{-1,1},\rho\in\brc{-1,1},\pi_1\in\brc{0,1}$, and $50\le m,n\le150$. Figure~\ref{fig:boxplot} presents the boxplots of the empirical TIEs obtained from these $10,000$ simulated scenarios for the likelihood ratio test ($Q_{LR}$), Wald-type test ($Q_W$) and score test ($Q_S$), respectively. We observe that the score test $Q_S$ exhibits a narrower interquartile range and a median closer to the nominal significance level ($\alpha=0.05$) compared to the other two tests. In contrast, the Wald-type test $Q_W$ shows a wider interquartile range and a median further from the nominal level compared to the other two tests. These results suggest that the score test $Q_S$ demonstrates slightly better stability and control of the type I error.

{\scriptsize
 \begin{longtblr}[
  caption={The empirical TIEs (in \%) of the three tests for $\rho=0$ under $H_0:~\delta=\delta_0$ at nominal level $\alpha=0.05$.},
   label={tab:TIE:1}
 ]{
   cell{1}{1}={r=2}{valign=m},
   cell{1}{2}={r=2}{valign=m},
   cell{1}{3}={r=2}{valign=m},
   cell{1}{4}={c=3}{halign=c},
   cell{1}{8}={c=3}{halign=c},
   cell{1}{12}={c=3}{halign=c},
   cell{3}{1}={r=9}{valign=h},
   cell{12}{1}={r=9}{valign=h},
   cell{21}{1}={r=9}{valign=h},
   cell{3}{2}={r=3}{valign=h},
   cell{6}{2}={r=3}{valign=h},
   cell{9}{2}={r=3}{valign=h},
   cell{12}{2}={r=3}{valign=h},
   cell{15}{2}={r=3}{valign=h},
   cell{18}{2}={r=3}{valign=h},
   cell{21}{2}={r=3}{valign=h},
   cell{24}{2}={r=3}{valign=h},
 }
 \toprule
 $\pi_1$ &$m$ &$n$ &$\delta_0=0.1$ & & & &$\delta_0=0.2$ & & & &$\delta_0=0.3$ & & \\
 \cline{4-6}\cline{8-10}\cline{12-14}
  & & &$Q_{LR}$ &$Q_W$ &$Q_S$ & &$Q_{LR}$ &$Q_W$ &$Q_S$ & &$Q_{LR}$ &$Q_W$ &$Q_S$ \\
 \midrule
0.1 & 50 & 50 &4.42 &4.63 &4.23 & &4.84 &5.13 &4.61 & &4.70 &5.05 &4.61 \\
& &100 &4.61 &4.71 &4.45 & &4.85 &5.03 &4.70 & &4.76 &4.98 &4.69 \\
& &150 &4.65 &4.72 &4.50 & &4.66 &4.85 &4.54 & &4.73 &4.92 &4.66 \\
&100 & 50 &4.79 &5.06 &4.61 & &4.64 &4.91 &4.52 & &4.73 &4.95 &4.65 \\
& &100 &4.65 &4.89 &4.53 & &4.72 &4.92 &4.62 & &4.74 &4.93 &4.66 \\
& &150 &4.78 &4.99 &4.66 & &4.70 &4.89 &4.63 & &4.80 &4.99 &4.74 \\
&150 & 50 &4.74 &4.96 &4.60 & &4.71 &4.90 &4.62 & &4.83 &4.95 &4.77 \\
& &100 &4.90 &5.12 &4.77 & &4.70 &4.88 &4.63 & &4.90 &5.01 &4.84 \\
& &150 &4.80 &4.97 &4.69 & &4.77 &4.92 &4.70 & &4.90 &5.01 &4.85 \\
0.2 & 50 & 50 &4.85 &5.07 &4.67 & &4.95 &5.22 &4.80 & &4.90 &5.16 &4.78 \\
& &100 &4.92 &5.13 &4.79 & &4.99 &5.15 &4.90 & &4.96 &5.12 &4.90 \\
& &150 &4.85 &4.99 &4.74 & &4.88 &5.03 &4.80 & &4.91 &5.02 &4.84 \\
&100 & 50 &5.04 &5.17 &4.92 & &5.15 &5.28 &5.06 & &5.03 &5.17 &4.95 \\
& &100 &5.02 &5.14 &4.93 & &5.04 &5.16 &4.96 & &5.04 &5.14 &4.98 \\
& &150 &5.03 &5.14 &4.94 & &5.06 &5.14 &4.99 & &5.08 &5.18 &5.03 \\
&150 & 50 &4.98 &5.08 &4.89 & &5.05 &5.14 &4.97 & &5.01 &5.13 &4.95 \\
& &100 &5.09 &5.16 &5.02 & &5.14 &5.24 &5.07 & &5.08 &5.15 &5.04 \\
& &150 &5.18 &5.25 &5.10 & &5.07 &5.14 &5.01 & &5.10 &5.18 &5.04 \\
0.3 & 50 & 50 &5.16 &5.39 &5.03 & &5.11 &5.42 &4.99 & &5.21 &5.43 &5.04 \\
& &100 &5.13 &5.29 &5.03 & &5.08 &5.21 &4.99 & &5.02 &5.24 &4.93 \\
& &150 &5.12 &5.25 &5.05 & &5.03 &5.15 &4.93 & &5.07 &5.13 &5.02 \\
&100 & 50 &5.23 &5.36 &5.15 & &5.15 &5.26 &5.06 & &5.21 &5.26 &5.16 \\
& &100 &5.05 &5.17 &4.97 & &5.05 &5.15 &4.98 & &5.01 &5.09 &4.97 \\
& &150 &5.12 &5.21 &5.07 & &5.14 &5.21 &5.11 & &5.15 &5.25 &5.08 \\
&150 & 50 &4.98 &5.07 &4.93 & &5.07 &5.12 &5.01 & &5.00 &5.10 &4.97 \\
& &100 &5.11 &5.20 &5.05 & &5.11 &5.18 &5.07 & &5.08 &5.20 &5.04 \\
& &150 &5.14 &5.21 &5.10 & &5.12 &5.22 &5.08 & &5.11 &5.13 &5.11 \\
 \bottomrule
 \end{longtblr}
}

{\scriptsize
 \begin{longtblr}[
  caption={The empirical TIEs (in \%) of the three tests for $\rho=0.5$ under $H_0:~\delta=\delta_0$ at nominal level $\alpha=0.05$.},
   label={tab:TIE:2}
 ]{
   cell{1}{1}={r=2}{valign=m},
   cell{1}{2}={r=2}{valign=m},
   cell{1}{3}={r=2}{valign=m},
   cell{1}{4}={c=3}{halign=c},
   cell{1}{8}={c=3}{halign=c},
   cell{1}{12}={c=3}{halign=c},
   cell{3}{1}={r=9}{valign=h},
   cell{12}{1}={r=9}{valign=h},
   cell{21}{1}={r=9}{valign=h},
   cell{3}{2}={r=3}{valign=h},
   cell{6}{2}={r=3}{valign=h},
   cell{9}{2}={r=3}{valign=h},
   cell{12}{2}={r=3}{valign=h},
   cell{15}{2}={r=3}{valign=h},
   cell{18}{2}={r=3}{valign=h},
   cell{21}{2}={r=3}{valign=h},
   cell{24}{2}={r=3}{valign=h},
 }
 \toprule
 $\pi_1$ &$m$ &$n$ &$\delta_0=0.1$ & & & &$\delta_0=0.2$ & & & &$\delta_0=0.3$ & & \\
 \cline{4-6}\cline{8-10}\cline{12-14}
  & & &$Q_{LR}$ &$Q_W$ &$Q_S$ & &$Q_{LR}$ &$Q_W$ &$Q_S$ & &$Q_{LR}$ &$Q_W$ &$Q_S$ \\
 \midrule
0.1 & 50 & 50 &5.10 &5.18 &4.94 & &5.10 &5.25 &4.97 & &5.04 &5.27 &4.92 \\
& &100 &5.01 &5.06 &4.91 & &5.03 &5.14 &4.95 & &5.00 &5.15 &4.93 \\
& &150 &5.03 &5.07 &4.95 & &4.93 &5.02 &4.87 & &4.99 &5.09 &4.94 \\
&100 & 50 &5.04 &5.10 &4.94 & &5.10 &5.17 &5.03 & &5.03 &5.13 &4.98 \\
& &100 &5.05 &5.09 &4.98 & &5.03 &5.13 &4.98 & &5.07 &5.16 &5.02 \\
& &150 &5.06 &5.08 &5.02 & &5.01 &5.08 &4.96 & &4.97 &5.04 &4.93 \\
&150 & 50 &4.98 &5.03 &4.93 & &5.05 &5.13 &4.99 & &4.99 &5.10 &4.94 \\
& &100 &5.13 &5.14 &5.07 & &5.00 &5.06 &4.95 & &5.04 &5.11 &5.01 \\
& &150 &5.07 &5.10 &5.01 & &5.07 &5.13 &5.04 & &5.06 &5.13 &5.04 \\
0.2 & 50 & 50 &5.06 &5.19 &4.98 & &5.06 &5.22 &4.94 & &5.07 &5.27 &5.01 \\
& &100 &5.13 &5.23 &5.07 & &5.07 &5.17 &5.01 & &5.07 &5.20 &5.04 \\
& &150 &5.03 &5.10 &4.99 & &4.92 &4.99 &4.91 & &5.02 &5.12 &5.00 \\
&100 & 50 &5.10 &5.19 &5.03 & &5.02 &5.12 &4.98 & &5.12 &5.25 &5.08 \\
& &100 &5.10 &5.18 &5.07 & &4.98 &5.06 &4.95 & &5.02 &5.19 &4.96 \\
& &150 &5.11 &5.18 &5.07 & &5.06 &5.14 &5.02 & &5.07 &5.15 &5.04 \\
&150 & 50 &5.00 &5.07 &4.96 & &4.90 &4.99 &4.87 & &4.98 &5.10 &4.95 \\
& &100 &5.08 &5.13 &5.04 & &5.09 &5.15 &5.07 & &5.12 &5.19 &5.08 \\
& &150 &5.08 &5.13 &5.04 & &5.10 &5.17 &5.08 & &5.17 &5.25 &5.11 \\
0.3 & 50 & 50 &5.12 &5.30 &5.03 & &5.10 &5.32 &5.02 & &5.20 &5.44 &5.13 \\
& &100 &5.08 &5.18 &5.01 & &5.02 &5.14 &4.99 & &5.07 &5.20 &5.03 \\
& &150 &5.05 &5.13 &5.01 & &5.00 &5.12 &4.98 & &4.99 &5.14 &4.93 \\
&100 & 50 &5.09 &5.21 &5.03 & &5.07 &5.20 &5.03 & &5.05 &5.20 &4.99 \\
& &100 &5.00 &5.10 &4.97 & &5.02 &5.13 &4.99 & &4.98 &5.09 &4.95 \\
& &150 &4.98 &5.07 &4.95 & &5.09 &5.17 &5.06 & &5.08 &5.19 &5.05 \\
&150 & 50 &5.04 &5.12 &4.99 & &5.03 &5.14 &5.00 & &5.03 &5.13 &4.99 \\
& &100 &5.05 &5.13 &5.01 & &5.10 &5.17 &5.07 & &5.11 &5.17 &5.06 \\
& &150 &5.09 &5.16 &5.07 & &5.07 &5.15 &5.05 & &5.06 &5.15 &5.05 \\
 \bottomrule
 \end{longtblr}
}

{\scriptsize
 \begin{longtblr}[
  caption={The empirical TIEs (in \%) of the three tests for $\rho=0.9$ under $H_0:~\delta=\delta_0$ at nominal level $\alpha=0.05$.},
   label={tab:TIE:3}
 ]{
   cell{1}{1}={r=2}{valign=m},
   cell{1}{2}={r=2}{valign=m},
   cell{1}{3}={r=2}{valign=m},
   cell{1}{4}={c=3}{halign=c},
   cell{1}{8}={c=3}{halign=c},
   cell{1}{12}={c=3}{halign=c},
   cell{3}{1}={r=9}{valign=h},
   cell{12}{1}={r=9}{valign=h},
   cell{21}{1}={r=9}{valign=h},
   cell{3}{2}={r=3}{valign=h},
   cell{6}{2}={r=3}{valign=h},
   cell{9}{2}={r=3}{valign=h},
   cell{12}{2}={r=3}{valign=h},
   cell{15}{2}={r=3}{valign=h},
   cell{18}{2}={r=3}{valign=h},
   cell{21}{2}={r=3}{valign=h},
   cell{24}{2}={r=3}{valign=h},
 }
 \toprule
 $\pi_1$ &$m$ &$n$ &$\delta_0=0.1$ & & & &$\delta_0=0.2$ & & & &$\delta_0=0.3$ & & \\
 \cline{4-6}\cline{8-10}\cline{12-14}
  & & &$Q_{LR}$ &$Q_W$ &$Q_S$ & &$Q_{LR}$ &$Q_W$ &$Q_S$ & &$Q_{LR}$ &$Q_W$ &$Q_S$ \\
 \midrule
0.1 & 50 & 50 &5.00 &5.07 &4.81 & &5.14 &5.29 &5.02 & &5.03 &5.23 &4.92 \\
& &100 &4.95 &5.04 &4.82 & &4.94 &5.04 &4.87 & &5.00 &5.12 &4.96 \\
& &150 &4.92 &4.96 &4.84 & &4.88 &4.97 &4.83 & &4.96 &5.06 &4.92 \\
&100 & 50 &5.03 &5.08 &4.92 & &4.98 &5.08 &4.92 & &5.01 &5.13 &4.95 \\
& &100 &4.96 &5.00 &4.85 & &4.95 &5.02 &4.87 & &5.00 &5.09 &4.96 \\
& &150 &5.02 &5.05 &4.95 & &5.05 &5.11 &4.98 & &4.98 &5.04 &4.94 \\
&150 & 50 &4.95 &5.00 &4.87 & &4.95 &5.03 &4.90 & &5.01 &5.12 &4.96 \\
& &100 &5.10 &5.14 &5.03 & &5.04 &5.11 &5.01 & &4.97 &5.04 &4.94 \\
& &150 &4.96 &4.97 &4.93 & &5.06 &5.10 &5.01 & &5.09 &5.14 &5.06 \\
0.2 & 50 & 50 &5.02 &5.17 &4.93 & &5.00 &5.19 &4.93 & &5.14 &5.34 &5.05 \\
& &100 &5.09 &5.20 &5.04 & &5.05 &5.18 &5.01 & &4.99 &5.13 &4.96 \\
& &150 &4.94 &5.02 &4.90 & &4.95 &5.02 &4.91 & &4.90 &5.00 &4.89 \\
&100 & 50 &5.15 &5.23 &5.07 & &5.10 &5.19 &5.05 & &4.99 &5.12 &4.97 \\
& &100 &4.95 &5.03 &4.90 & &5.00 &5.06 &4.97 & &4.89 &5.01 &4.85 \\
& &150 &5.01 &5.08 &4.97 & &5.09 &5.16 &5.07 & &5.05 &5.11 &5.03 \\
&150 & 50 &4.98 &5.04 &4.93 & &5.01 &5.09 &4.98 & &4.96 &5.08 &4.93 \\
& &100 &5.02 &5.07 &4.98 & &5.06 &5.15 &5.06 & &5.01 &5.08 &5.00 \\
& &150 &5.07 &5.11 &5.04 & &5.09 &5.15 &5.08 & &5.12 &5.20 &5.10 \\
0.3 & 50 & 50 &5.07 &5.22 &5.01 & &5.15 &5.33 &5.13 & &5.16 &5.42 &5.07 \\
& &100 &5.08 &5.18 &5.04 & &5.02 &5.14 &4.98 & &5.05 &5.19 &5.01 \\
& &150 &4.98 &5.07 &4.95 & &5.00 &5.10 &4.96 & &4.98 &5.07 &4.94 \\
&100 & 50 &5.06 &5.16 &5.02 & &4.99 &5.13 &4.96 & &5.01 &5.21 &4.97 \\
& &100 &4.97 &5.05 &4.95 & &4.98 &5.10 &4.96 & &4.97 &5.07 &4.93 \\
& &150 &5.09 &5.16 &5.08 & &5.07 &5.14 &5.04 & &5.06 &5.13 &5.05 \\
&150 & 50 &4.96 &5.04 &4.94 & &4.97 &5.05 &4.93 & &4.97 &5.10 &4.96 \\
& &100 &5.13 &5.20 &5.12 & &5.04 &5.10 &5.03 & &5.08 &5.15 &5.01 \\
& &150 &5.09 &5.15 &5.06 & &5.14 &5.23 &5.12 & &5.14 &5.20 &5.10 \\
 \bottomrule
 \end{longtblr}
}

\subsubsection{Empirical Powers}
\label{sec:results:simulation:power}
The empirical powers are computed in the same manner as the empirical TIEs, except that the datasets are generated under the alternative hypothesis $H_1:~\delta=\delta_1$ with $\delta_1= 0.1,0.2,0.3$, against the null hypothesis $H_0:~\delta=0$. Tables~\ref{tab:power:1} - \ref{tab:power:3} present the empirical power results at the nominal significance level $\alpha=0.05$ for $\rho=0$, $0.5$ and $0.9$, respectively. The results show that the empirical powers of the three tests are generally comparable under the same parameter settings. Moreover, for each given value of $\rho$, the empirical powers increase with the sample size and risk difference in proportions $\delta_1$ (hereafter referred to as the \textit{effect size}). On the other hand, for fixed sample size and effect size, the empirical powers decrease as the intra-subject correlation $\rho$ increases.  

{\scriptsize
 \begin{longtblr}[
  caption={The empirical powers (in \%) of the three tests for $\rho=0$ under $H_0:~\delta=0$ vs $H_1:~\delta=\delta_1$ at nominal level $\alpha=0.05$.},
   label={tab:power:1}
 ]{
   cell{1}{1}={r=2}{valign=m},
   cell{1}{2}={r=2}{valign=m},
   cell{1}{3}={r=2}{valign=m},
   cell{1}{4}={c=3}{halign=c},
   cell{1}{8}={c=3}{halign=c},
   cell{1}{12}={c=3}{halign=c},
   cell{3}{1}={r=9}{valign=h},
   cell{12}{1}={r=9}{valign=h},
   cell{21}{1}={r=9}{valign=h},
   cell{3}{2}={r=3}{valign=h},
   cell{6}{2}={r=3}{valign=h},
   cell{9}{2}={r=3}{valign=h},
   cell{12}{2}={r=3}{valign=h},
   cell{15}{2}={r=3}{valign=h},
   cell{18}{2}={r=3}{valign=h},
   cell{21}{2}={r=3}{valign=h},
   cell{24}{2}={r=3}{valign=h},
 }
 \toprule
 $\pi_1$ &$m$ &$n$ &$\delta_1=0.1$ & & & &$\delta_1=0.2$ & & & &$\delta_1=0.3$ & & \\
 \cline{4-6}\cline{8-10}\cline{12-14}
  & & &$Q_{LR}$ &$Q_W$ &$Q_S$ & &$Q_{LR}$ &$Q_W$ &$Q_S$ & &$Q_{LR}$ &$Q_W$ &$Q_S$ \\
 \midrule
0.1 & 50 & 50 & 65.23 & 65.52 & 64.51 & & 99.26 & 99.28 & 99.21 & &100.00 &100.00 &100.00 \\
& &100 & 78.69 & 78.88 & 78.32 & & 99.92 & 99.92 & 99.92 & &100.00 &100.00 &100.00 \\
& &150 & 87.65 & 87.73 & 87.48 & & 99.99 & 99.99 & 99.99 & &100.00 &100.00 &100.00 \\
&100 & 50 & 87.86 & 88.08 & 87.52 & & 99.99 & 99.99 & 99.99 & &100.00 &100.00 &100.00 \\
& &100 & 93.07 & 93.20 & 92.90 & &100.00 &100.00 &100.00 & &100.00 &100.00 &100.00 \\
& &150 & 96.13 & 96.18 & 96.04 & &100.00 &100.00 &100.00 & &100.00 &100.00 &100.00 \\
&150 & 50 & 96.28 & 96.36 & 96.17 & &100.00 &100.00 &100.00 & &100.00 &100.00 &100.00 \\
& &100 & 97.91 & 97.95 & 97.83 & &100.00 &100.00 &100.00 & &100.00 &100.00 &100.00 \\
& &150 & 98.89 & 98.91 & 98.85 & &100.00 &100.00 &100.00 & &100.00 &100.00 &100.00 \\
0.2 & 50 & 50 & 51.39 & 52.14 & 50.74 & & 96.89 & 97.02 & 96.81 & & 99.99 & 99.99 & 99.98 \\
& &100 & 63.72 & 64.27 & 63.23 & & 99.30 & 99.34 & 99.28 & &100.00 &100.00 &100.00 \\
& &150 & 73.52 & 73.89 & 73.18 & & 99.84 & 99.85 & 99.84 & &100.00 &100.00 &100.00 \\
&100 & 50 & 73.84 & 74.27 & 73.49 & & 99.84 & 99.85 & 99.84 & &100.00 &100.00 &100.00 \\
& &100 & 81.40 & 81.66 & 81.19 & & 99.97 & 99.97 & 99.97 & &100.00 &100.00 &100.00 \\
& &150 & 86.33 & 86.52 & 86.19 & & 99.99 & 99.99 & 99.99 & &100.00 &100.00 &100.00 \\
&150 & 50 & 86.81 & 87.01 & 86.65 & & 99.99 & 99.99 & 99.99 & &100.00 &100.00 &100.00 \\
& &100 & 90.61 & 90.73 & 90.50 & &100.00 &100.00 &100.00 & &100.00 &100.00 &100.00 \\
& &150 & 93.66 & 93.73 & 93.59 & &100.00 &100.00 &100.00 & &100.00 &100.00 &100.00 \\
0.3 & 50 & 50 & 44.65 & 45.35 & 44.20 & & 94.62 & 94.80 & 94.51 & & 99.97 & 99.97 & 99.96 \\
& &100 & 55.79 & 56.32 & 55.46 & & 98.46 & 98.51 & 98.43 & &100.00 &100.00 &100.00 \\
& &150 & 65.51 & 65.93 & 65.23 & & 99.57 & 99.58 & 99.56 & &100.00 &100.00 &100.00 \\
&100 & 50 & 65.25 & 65.66 & 65.04 & & 99.59 & 99.59 & 99.58 & &100.00 &100.00 &100.00 \\
& &100 & 73.16 & 73.43 & 72.99 & & 99.90 & 99.90 & 99.90 & &100.00 &100.00 &100.00 \\
& &150 & 79.31 & 79.51 & 79.17 & & 99.97 & 99.98 & 99.97 & &100.00 &100.00 &100.00 \\
&150 & 50 & 79.61 & 79.82 & 79.50 & & 99.97 & 99.97 & 99.97 & &100.00 &100.00 &100.00 \\
& &100 & 84.40 & 84.56 & 84.31 & &100.00 &100.00 &100.00 & &100.00 &100.00 &100.00 \\
& &150 & 88.39 & 88.50 & 88.30 & &100.00 &100.00 &100.00 & &100.00 &100.00 &100.00 \\
 \bottomrule
 \end{longtblr}
}

{\scriptsize
 \begin{longtblr}[
  caption={The empirical powers (in \%) of the three tests for $\rho=0.5$ under $H_0:~\delta=0$ vs $H_1:~\delta=\delta_1$ at nominal level $\alpha=0.05$.},
   label={tab:power:2}
 ]{
   cell{1}{1}={r=2}{valign=m},
   cell{1}{2}={r=2}{valign=m},
   cell{1}{3}={r=2}{valign=m},
   cell{1}{4}={c=3}{halign=c},
   cell{1}{8}={c=3}{halign=c},
   cell{1}{12}={c=3}{halign=c},
   cell{3}{1}={r=9}{valign=h},
   cell{12}{1}={r=9}{valign=h},
   cell{21}{1}={r=9}{valign=h},
   cell{3}{2}={r=3}{valign=h},
   cell{6}{2}={r=3}{valign=h},
   cell{9}{2}={r=3}{valign=h},
   cell{12}{2}={r=3}{valign=h},
   cell{15}{2}={r=3}{valign=h},
   cell{18}{2}={r=3}{valign=h},
   cell{21}{2}={r=3}{valign=h},
   cell{24}{2}={r=3}{valign=h},
 }
 \toprule
 $\pi_1$ &$m$ &$n$ &$\delta_1=0.1$ & & & &$\delta_1=0.2$ & & & &$\delta_1=0.3$ & & \\
 \cline{4-6}\cline{8-10}\cline{12-14}
  & & &$Q_{LR}$ &$Q_W$ &$Q_S$ & &$Q_{LR}$ &$Q_W$ &$Q_S$ & &$Q_{LR}$ &$Q_W$ &$Q_S$ \\
 \midrule
0.1 & 50 & 50 & 59.63 & 59.67 & 59.08 & & 97.89 & 97.94 & 97.83 & & 99.98 & 99.98 & 99.98 \\
& &100 & 74.54 & 74.60 & 74.24 & & 99.74 & 99.74 & 99.73 & &100.00 &100.00 &100.00 \\
& &150 & 84.90 & 84.94 & 84.72 & & 99.98 & 99.98 & 99.98 & &100.00 &100.00 &100.00 \\
&100 & 50 & 79.36 & 79.37 & 79.11 & & 99.89 & 99.89 & 99.88 & &100.00 &100.00 &100.00 \\
& &100 & 87.79 & 87.81 & 87.65 & & 99.99 & 99.99 & 99.99 & &100.00 &100.00 &100.00 \\
& &150 & 92.93 & 92.94 & 92.85 & &100.00 &100.00 &100.00 & &100.00 &100.00 &100.00 \\
&150 & 50 & 90.24 & 90.23 & 90.12 & & 99.99 & 99.99 & 99.99 & &100.00 &100.00 &100.00 \\
& &100 & 94.34 & 94.33 & 94.28 & &100.00 &100.00 &100.00 & &100.00 &100.00 &100.00 \\
& &150 & 96.91 & 96.91 & 96.89 & &100.00 &100.00 &100.00 & &100.00 &100.00 &100.00 \\
0.2 & 50 & 50 & 43.31 & 43.74 & 43.00 & & 92.49 & 92.68 & 92.40 & & 99.85 & 99.86 & 99.85 \\
& &100 & 56.93 & 57.22 & 56.69 & & 98.22 & 98.27 & 98.20 & &100.00 &100.00 &100.00 \\
& &150 & 68.21 & 68.43 & 68.06 & & 99.62 & 99.63 & 99.62 & &100.00 &100.00 &100.00 \\
&100 & 50 & 61.49 & 61.74 & 61.31 & & 98.98 & 99.01 & 98.98 & &100.00 &100.00 &100.00 \\
& &100 & 71.68 & 71.86 & 71.56 & & 99.78 & 99.78 & 99.78 & &100.00 &100.00 &100.00 \\
& &150 & 79.23 & 79.35 & 79.13 & & 99.96 & 99.96 & 99.96 & &100.00 &100.00 &100.00 \\
&150 & 50 & 74.86 & 75.03 & 74.75 & & 99.88 & 99.89 & 99.88 & &100.00 &100.00 &100.00 \\
& &100 & 81.70 & 81.81 & 81.61 & & 99.98 & 99.98 & 99.98 & &100.00 &100.00 &100.00 \\
& &150 & 87.18 & 87.26 & 87.13 & &100.00 &100.00 &100.00 & &100.00 &100.00 &100.00 \\
0.3 & 50 & 50 & 36.75 & 37.32 & 36.53 & & 88.37 & 88.66 & 88.26 & & 99.71 & 99.72 & 99.70 \\
& &100 & 48.73 & 49.10 & 48.57 & & 96.44 & 96.52 & 96.40 & & 99.99 & 99.99 & 99.99 \\
& &150 & 59.43 & 59.71 & 59.31 & & 99.06 & 99.08 & 99.06 & &100.00 &100.00 &100.00 \\
&100 & 50 & 52.55 & 52.94 & 52.41 & & 97.67 & 97.73 & 97.65 & & 99.99 & 99.99 & 99.99 \\
& &100 & 62.64 & 62.89 & 62.52 & & 99.39 & 99.40 & 99.38 & &100.00 &100.00 &100.00 \\
& &150 & 70.71 & 70.90 & 70.64 & & 99.85 & 99.86 & 99.85 & &100.00 &100.00 &100.00 \\
&150 & 50 & 65.88 & 66.13 & 65.79 & & 99.59 & 99.59 & 99.58 & &100.00 &100.00 &100.00 \\
& &100 & 73.29 & 73.50 & 73.21 & & 99.91 & 99.91 & 99.91 & &100.00 &100.00 &100.00 \\
& &150 & 79.67 & 79.80 & 79.61 & & 99.98 & 99.98 & 99.98 & &100.00 &100.00 &100.00 \\
 \bottomrule
 \end{longtblr}
}

{\scriptsize
 \begin{longtblr}[
  caption={The empirical powers (in \%) of the three tests for $\rho=0.9$ under $H_0:~\delta=0$ vs $H_1:~\delta=\delta_1$ at nominal level $\alpha=0.05$.},
   label={tab:power:3}
 ]{
   cell{1}{1}={r=2}{valign=m},
   cell{1}{2}={r=2}{valign=m},
   cell{1}{3}={r=2}{valign=m},
   cell{1}{4}={c=3}{halign=c},
   cell{1}{8}={c=3}{halign=c},
   cell{1}{12}={c=3}{halign=c},
   cell{3}{1}={r=9}{valign=h},
   cell{12}{1}={r=9}{valign=h},
   cell{21}{1}={r=9}{valign=h},
   cell{3}{2}={r=3}{valign=h},
   cell{6}{2}={r=3}{valign=h},
   cell{9}{2}={r=3}{valign=h},
   cell{12}{2}={r=3}{valign=h},
   cell{15}{2}={r=3}{valign=h},
   cell{18}{2}={r=3}{valign=h},
   cell{21}{2}={r=3}{valign=h},
   cell{24}{2}={r=3}{valign=h},
 }
 \toprule
 $\pi_1$ &$m$ &$n$ &$\delta_1=0.1$ & & & &$\delta_1=0.2$ & & & &$\delta_1=0.3$ & & \\
 \cline{4-6}\cline{8-10}\cline{12-14}
  & & &$Q_{LR}$ &$Q_W$ &$Q_S$ & &$Q_{LR}$ &$Q_W$ &$Q_S$ & &$Q_{LR}$ &$Q_W$ &$Q_S$ \\
 \midrule
0.1 & 50 & 50 & 52.92 & 52.92 & 52.29 & & 95.90 & 95.94 & 95.80 & & 99.95 & 99.96 & 99.95 \\
& &100 & 69.92 & 69.96 & 69.57 & & 99.46 & 99.47 & 99.45 & &100.00 &100.00 &100.00 \\
& &150 & 81.85 & 81.88 & 81.63 & & 99.95 & 99.95 & 99.95 & &100.00 &100.00 &100.00 \\
&100 & 50 & 71.20 & 71.23 & 70.85 & & 99.53 & 99.54 & 99.52 & &100.00 &100.00 &100.00 \\
& &100 & 82.58 & 82.61 & 82.41 & & 99.96 & 99.96 & 99.96 & &100.00 &100.00 &100.00 \\
& &150 & 89.68 & 89.69 & 89.55 & & 99.99 &100.00 & 99.99 & &100.00 &100.00 &100.00 \\
&150 & 50 & 83.38 & 83.39 & 83.18 & & 99.96 & 99.96 & 99.96 & &100.00 &100.00 &100.00 \\
& &100 & 90.13 & 90.15 & 90.03 & &100.00 &100.00 &100.00 & &100.00 &100.00 &100.00 \\
& &150 & 94.44 & 94.44 & 94.38 & &100.00 &100.00 &100.00 & &100.00 &100.00 &100.00 \\
0.2 & 50 & 50 & 38.62 & 39.05 & 38.34 & & 88.78 & 89.04 & 88.67 & & 99.62 & 99.63 & 99.62 \\
& &100 & 52.71 & 53.00 & 52.52 & & 97.21 & 97.27 & 97.19 & & 99.99 & 99.99 & 99.99 \\
& &150 & 64.95 & 65.16 & 64.82 & & 99.39 & 99.41 & 99.39 & &100.00 &100.00 &100.00 \\
&100 & 50 & 53.56 & 53.84 & 53.37 & & 97.46 & 97.50 & 97.44 & & 99.99 & 99.99 & 99.99 \\
& &100 & 65.68 & 65.88 & 65.54 & & 99.44 & 99.46 & 99.44 & &100.00 &100.00 &100.00 \\
& &150 & 74.57 & 74.70 & 74.50 & & 99.91 & 99.91 & 99.91 & &100.00 &100.00 &100.00 \\
&150 & 50 & 66.40 & 66.60 & 66.25 & & 99.51 & 99.52 & 99.51 & &100.00 &100.00 &100.00 \\
& &100 & 75.14 & 75.27 & 75.06 & & 99.92 & 99.92 & 99.92 & &100.00 &100.00 &100.00 \\
& &150 & 82.36 & 82.46 & 82.31 & & 99.98 & 99.98 & 99.98 & &100.00 &100.00 &100.00 \\
0.3 & 50 & 50 & 32.82 & 33.34 & 32.66 & & 84.13 & 84.52 & 84.03 & & 99.29 & 99.32 & 99.28 \\
& &100 & 45.38 & 45.71 & 45.25 & & 95.03 & 95.14 & 95.00 & & 99.97 & 99.98 & 99.97 \\
& &150 & 56.38 & 56.68 & 56.27 & & 98.61 & 98.63 & 98.60 & &100.00 &100.00 &100.00 \\
&100 & 50 & 46.09 & 46.44 & 45.96 & & 95.41 & 95.52 & 95.39 & & 99.98 & 99.98 & 99.98 \\
& &100 & 57.02 & 57.30 & 56.94 & & 98.72 & 98.75 & 98.71 & &100.00 &100.00 &100.00 \\
& &150 & 65.94 & 66.15 & 65.87 & & 99.66 & 99.67 & 99.66 & &100.00 &100.00 &100.00 \\
&150 & 50 & 57.65 & 57.92 & 57.57 & & 98.82 & 98.84 & 98.81 & &100.00 &100.00 &100.00 \\
& &100 & 66.41 & 66.62 & 66.34 & & 99.69 & 99.69 & 99.69 & &100.00 &100.00 &100.00 \\
& &150 & 74.19 & 74.33 & 74.14 & & 99.92 & 99.92 & 99.92 & &100.00 &100.00 &100.00 \\
 \bottomrule
 \end{longtblr}
}

Table~\ref{tab:power:size} presents the mimimal sample sizes (assuming $m=n$) required to achieve $80\%$ power for given values of $\rho$, $\pi_1$ and the effect size $\delta_1$ for the three tests. Based on this table, one may use interpolation to determine the minimal sample size needed to attain $80\%$ power for other combinations of $\rho$, $\pi_1$ and $\delta_1$.  

{\scriptsize
 \begin{longtblr}[
   caption={Sample size with power = 80\%},
   label={tab:power:size}
 ]{
   cell{1}{1}={r=2}{valign=m},
   cell{1}{2}={r=2}{valign=m},
   cell{1}{3}={c=3}{halign=c},
   cell{1}{7}={c=3}{halign=c},
   cell{1}{11}={c=3}{halign=c},
   cell{3}{1}={r=5}{valign=h},
   cell{8}{1}={r=5}{valign=h},
   cell{13}{1}={r=5}{valign=h},
   cell{18}{1}={r=5}{valign=h},
   cell{23}{1}={r=5}{valign=h},
 }
 \toprule
 $\rho$ &$\pi_1$ &$\delta_1=0.1$ & & & &$\delta_1=0.2$ & & & &$\delta_1=0.3$ & & \\
 \cline{3-5}\cline{7-9}\cline{11-13}
   & &$Q_{LR}$ &$Q_W$ &$Q_S$ & &$Q_{LR}$ &$Q_W$ &$Q_S$ & &$Q_{LR}$ &$Q_W$ &$Q_S$ \\
 \midrule
0.0 &0.1 & 68 & 69 & 69 & & 23 & 23 & 23 & & 12 & 12 & 13 \\
    &0.2 & 97 & 98 & 97 & & 28 & 28 & 29 & & 14 & 14 & 14 \\
    &0.3 &118 &116 &118 & & 31 & 31 & 32 & & 15 & 14 & 15 \\
    &0.4 &129 &127 &132 & & 33 & 32 & 33 & & 15 & 14 & 15 \\
    &0.5 &129 &130 &130 & & 31 & 31 & 32 & & 14 & 14 & 14 \\
0.2 &0.1 & 73 & 72 & 73 & & 24 & 23 & 24 & & 13 & 13 & 13 \\
    &0.2 &108 &108 &111 & & 30 & 30 & 30 & & 15 & 14 & 15 \\
    &0.3 &132 &132 &132 & & 35 & 34 & 36 & & 16 & 15 & 16 \\
    &0.4 &145 &146 &145 & & 37 & 36 & 37 & & 16 & 16 & 16 \\
    &0.5 &144 &144 &144 & & 36 & 35 & 36 & & 15 & 15 & 15 \\
0.4 &0.1 & 78 & 78 & 78 & & 25 & 24 & 25 & & 13 & 13 & 14 \\
    &0.2 &120 &118 &119 & & 33 & 33 & 33 & & 16 & 16 & 16 \\
    &0.3 &146 &145 &148 & & 39 & 38 & 38 & & 17 & 17 & 18 \\
    &0.4 &161 &158 &159 & & 40 & 39 & 40 & & 18 & 17 & 18 \\
    &0.5 &160 &161 &159 & & 38 & 39 & 38 & & 16 & 16 & 16 \\
0.6 &0.1 & 82 & 83 & 85 & & 26 & 26 & 27 & & 14 & 14 & 15 \\
    &0.2 &128 &128 &129 & & 35 & 35 & 35 & & 17 & 17 & 17 \\
    &0.3 &160 &155 &157 & & 41 & 41 & 42 & & 19 & 18 & 19 \\
    &0.4 &173 &173 &174 & & 43 & 43 & 44 & & 19 & 19 & 19 \\
    &0.5 &174 &172 &174 & & 41 & 41 & 42 & & 17 & 17 & 18 \\
0.8 &0.1 & 91 & 90 & 92 & & 29 & 29 & 28 & & 15 & 15 & 15 \\
    &0.2 &135 &133 &137 & & 38 & 38 & 39 & & 18 & 18 & 19 \\
    &0.3 &167 &165 &167 & & 44 & 44 & 43 & & 20 & 20 & 20 \\
    &0.4 &183 &183 &186 & & 47 & 45 & 46 & & 20 & 20 & 20 \\
    &0.5 &185 &184 &182 & & 44 & 44 & 45 & & 18 & 18 & 18 \\
 \bottomrule
 \end{longtblr}
}

To further assist researchers in determining the sample size (without assuming $m=n$), power and effect size under arbitrary parameter settings, we have developed a user-friendly online calculator available at~\href{https://www.buffalo.edu/~cxma/test_RiskDiffRhoModelCombined.htm}{https://www.buffalo.edu/~cxma/test\_RiskDiffRhoModelCombined.htm}.

Based on the simulation studies, we conclude that the score test performs slightly better than the likelihood ratio and Wald-type tests, owing to its marginally better control of the type I error and comparable power. It is therefore recommended for practical applications in future studies.

\subsection{Real-World Example}
\label{sec:results:real}
Two real-world examples are presented to illustrate the test for the risk difference between two proportions under $H_0:~\delta = 0$, which is equivalent to testing equal proportions in two groups.

The first example involves a subset of 214 children who were admitted with acute otitis media with effusion (OME) and randomized into two treatment groups receiving cefaclor or amoxicillin, respectively~\cite{mandel1982duration}. Table~\ref{tab:OME} presents the number of cured ears among 173 children evaluated at 42 days.
\begin{table}[thpb]
    \centering
    \caption{Number of cured ears at 42 days in children treated with cefaclor and amoxicillin.}
    \label{tab:OME}
    \begin{tabular}{cccc}
    \toprule
         &\multicolumn{2}{c}{Treatment} & \\
         \cmidrule{2-3}
        \# of cured ears &Cefaclor &Amoxicillin &total \\
        \midrule
        0 &9 &7 &16 \\
        1 &7 &5 &12 \\
        2 &23 &13 &36 \\
        total &39 &25 &64 \\
        \midrule
        0 &20 &19 &39 \\
        1 &34 &36 &70 \\
        total &54 &55 &109 \\
        \bottomrule
    \end{tabular}
\end{table}

Goodness-of-fit tests are performed for the OME dataset under Donner's model, using the five recommended methos (deviance test ($G^2$), Pearson chi-square test ($X^2$), and three boostrap methods ($B_1,B_2,B_3$)) by Zhou and Ma~\cite{zhou2025goodness}. All $p$-values from these methods are greater than $0.9$, indicating that Donner's model provides an adequate fit for the OME dataset. 
Table~\ref{tab:OME:stat} presents the values of the three test statistics and their corresponding $p$-values, along with the constrained and unconstrained MLEs, for the OME dataset. The $p$-values of the three test statistics are very close to each other and appear identical when rounded to four decimal places, all being much larger than the nominal significance level $\alpha=0.05$. Therefore, we fail to reject the null hypothesis $H_0:~\delta=0$ and conclude that there is no significant difference in the curing proportions of infected ears between cefaclor and amoxicillin treatments. 
\begin{table}[thpb]
  \centering
  \caption{Example 1: statistic and $p$-values along with constrained and unconstrained MLEs.}
  \label{tab:OME:stat}
  \begin{tabular}{cccccc}
    \toprule
    \multicolumn{2}{c}{Constrained MLEs} & &\multicolumn{3}{c}{Unconstrained MLEs} \\
    \cmidrule{1-2} \cmidrule{4-6}
    $\hat{\pi}_{1,0}$ &$\hat{\rho}_0$ & &$\hat{\delta}$ &$\hat{\pi}_1$ &$\hat{\rho}$ \\
    \midrule
    0.6482 &0.5862 & &-0.0119 &0.6536 &0.5856 \\
    \midrule
    \multicolumn{6}{c}{Test Statistics} \\
    \midrule
    & & &$Q_{LR}$ &$Q_W$ &$Q_S$ \\
    \midrule
    \multicolumn{2}{c}{statistics} & &0.0293 &0.0293 &0.0293 \\
    \multicolumn{2}{c}{$p$-value} & &0.8641 &0.8641 &0.8641 \\
    \bottomrule
  \end{tabular}
\end{table}

The second example consists of combined unilateral and bilateral data obtained from an observational study for sixty myopia patients receiving the so-called Orthokeratology (Ortho-k), a non-surgical vision correction method that uses specialized contact lenses worn overnight to temporarily reshape the cornea and correct myopia \cite{liang2024homogeneity}. There are two lens designs regarding Orth-k treatment method: (i) corneal refractive therapy (CRT) used by brand S; (ii) vision shaping treatment (VST) used by other brands. Myopia improvement is assessed by the axial length growth (ALG), where improvement is indicated if ALG is less than $0.3$ mm, and absent otherwise. The observations on the number of improved myopic eyes by lens designs are summarized in Table \ref{tab:myopia}.
\begin{table}[thpb]
    \centering
    \caption{Number of improved myopic eyes with Ortho-k treatment using VST and CRT lens designs.}
    \label{tab:myopia}
    \begin{tabular}{cccc}
    \hline
         &\multicolumn{2}{c}{Lens Design} & \\
         \cline{2-3}
        \# of myopia improved eyes &VST &CRT &total \\
        \hline
        0 &20 &13 &33 \\
        1 &7 &2 &9 \\
        2 &10 &2 &12 \\
        total &37 &17 &54 \\
        \hline
        0 &3 &0 &3 \\
        1 &3 &0 &3 \\
        total &6 &0 &6 \\
        \hline
    \end{tabular}
\end{table}

The same goodness-of-fit tests are perform to the Ortho-K dataset under Donner's model, where all $p$-values from the five recommended methods are greater than $0.9$, suggesting that Donner's model provides an adequate fit for the Orhto-K dataset as well. 
Table~\ref{tab:myopia:stat} presents the values of the three test statistics and their corresponding $p$-values, along with the constrained and unconstrained MLEs, for the Ortho-k dataset. As can be seen, all the $p$-values for the three tests are greater than the nominal significance level $\alpha=0.05$. Therefore, we fail to reject the null hypothesis and conclude that there is no significant difference in the proportions of improved myopic eyes between the two lens designs. 
\begin{table}[thpb]
  \centering
  \caption{Example 2: statistic and $p$-values along with constrained and unconstrained MLEs.}
  \label{tab:myopia:stat}
  \begin{tabular}{cccccc}
    \toprule
    \multicolumn{2}{c}{Constrained MLEs} & &\multicolumn{3}{c}{Unconstrained MLEs} \\
    \cmidrule{1-2} \cmidrule{4-6}
    $\hat{\pi}_{1,0}$ &$\hat{\rho}_0$ & &$\hat{\delta}$ &$\hat{\pi}_1$ &$\hat{\rho}$ \\
    \midrule
    0.3215 &0.6117 & &-0.2039 &0.3803 &0.5948 \\
    \midrule
    \multicolumn{6}{c}{Test Statistics} \\
    \midrule
    & & &$Q_{LR}$ &$Q_W$ &$Q_S$ \\
    \midrule
    \multicolumn{2}{c}{statistics} & &3.1689 &3.7843 &2.9671 \\
    \multicolumn{2}{c}{$p$-value} & &0.0751 &0.0517 &0.0850 \\
    \bottomrule
  \end{tabular}
\end{table}

To facilitate wider application, we have extended the functionality of the~\href{https://www.buffalo.edu/~cxma/test_RiskDiffRhoModelCombined.htm}{online calculator} provided previously to allow users to perform the proposed risk difference test using their own datasets.

\section{Conclusions}
\label{sec:conclusions}
In this study, we investigated three likelihood-based test statistics: the likelihood ratio test ($Q_{LR}$), Wald-typ test ($Q_W$) and score test ($Q_S$), respectively, for testing the risk difference of two proportions for combined unilateral and bilateral data. Through extensive simulation studies, we evaluated their empirical TIEs and powers under various parameter settings. The results show that all three tests maintain fairly good control of the type I error across a wide range of scenarios. Among them, the score test consistently demonstrates slightly better perfomrance, showing narrower variability in empirical TIEs and power estimates that are comparable to those of the other two tests.

Power analysis further revealed that, as expected, power increases with sample size and effect size (risk difference), but decreases with higher intra-subject correlation. A useful~\href{https://www.buffalo.edu/~cxma/test_RiskDiffRhoModelCombined.htm}{online calculator} has been developed to assist researchers in determining sample sizes, power, and effect sizes under arbitrary parameter settings.

Two real examples from otolaryngologic and ophthalmologic studies were analyzed to illustrate the application of the proposed methods. In addition, the~\href{https://www.buffalo.edu/~cxma/test_RiskDiffRhoModelCombined.htm}{online calculator} enables users to perform the proposed risk difference test on their own datasets. 

Overall, the score test provides marginally superior control of type I error and comparable power, making it a recommended choice for practical applications and future studies involving combined unilateral and bilateral data.

Finally, we note that the proposed statistics rely on large sample asymptotic properties and thus perform best with sufficiently large sample sizes. For small sample situations, exact methods may be more appropriate and will be explored in future work.

\appendix
\section{Fisher Information Matrix}\label{app:FI}
The Fisher information matrix is defined as the variance-covariance matrix of the score. Under certain regularity conditions, it can be written as 
\begin{equation}
  \bI\brc{\bbeta}=\mathbb{E}\brc{-\frac{\partial^2\ell\brc{\bbeta}}{\partial\bbeta^T\partial\bbeta}}, 
\end{equation}
with the parameter vector $\bbeta$ of interest and its corresponding log-likelihood function $\ell\brc{\bbeta}$.
Let $\bbeta=\bgam=\brc{\delta,\pi_1,\rho}^T$ and $\ell\brc{\bbeta}=\ell_1\brc{\bgam}$ in equation (\ref{eq:loglik:2}), we have
\begin{equation}
  \bI\brc{\bgam}=
  \bI\brc{\delta,\pi_1,\rho}=
  \mathbb{E}
  \brc{
    \begin{array}{ccc}
      -\frac{\partial^2\ell_1}{\partial\delta^2} &-\frac{\partial^2\ell_1}{\partial\delta\partial\pi_1} &-\frac{\partial^2\ell_1}{\partial\delta\partial\rho} \\
      -\frac{\partial^2\ell_1}{\partial\pi_1\partial\delta} &-\frac{\partial^2\ell_1}{\partial\pi^2} &-\frac{\partial^2\ell_1}{\partial\pi_1\partial\rho} \\
      -\frac{\partial^2\ell_1}{\partial\rho\partial\delta} &-\frac{\partial^2\ell_1}{\partial\rho\partial\pi_1} &-\frac{\partial^2\ell_1}{\partial\rho^2}
    \end{array}
  }
  =
  \brc{
    \begin{array}{ccc}
      I_{11} &I_{12} &I_{13} \\
      I_{21} &I_{22} &I_{23} \\
      I_{31} &I_{32} &I_{33}
    \end{array}
  }, 
\end{equation}
where the elements $I_{ij}$ ($i,j=1,2,3$) are given by
\begin{align*}
  I_{11}=&\frac{m_{+2}\brc{2-\rho}+n_{+2}}{\brc{\delta+\pi_1}\brc{1-\delta-\pi_1}}
  -\frac{m_{+2}\rho\brc{1-\rho^2}}{\brc{1-\brc{1-\rho}\brc{\delta+\pi_1}}\brc{\rho+\brc{1-\rho}\brc{\delta+\pi_1}}}, \\
  I_{12}=&I_{11}, \\
  I_{13}=&\frac{m_{+2}\rho\brc{2\brc{\delta+\pi_1}-1}}{\brc{1-\brc{1-\rho}\brc{\delta+\pi_1}}\brc{\rho+\brc{1-\rho}\brc{\delta+\pi_1}}}, \\
  I_{21}=&I_{12}, \\
  I_{22}=&\frac{m_{+1}\brc{2-\rho}+n_{+1}}{\pi_1\brc{1-\pi_1}}+\frac{m_{+2}\brc{2-\rho}+n_{+2}}{\brc{1-\delta-\pi_1}\brc{\delta+\pi_1}}+\frac{m_{+1}\rho\brc{1-\rho^2}}{\brc{1-\brc{1-\rho}\pi_1}\brc{\rho+\brc{1-\rho}\pi_1}} \\
  &-\frac{m_{+2}\rho\brc{1-\rho^2}}{\brc{1-\brc{1-\rho}\brc{\delta+\pi_1}}\brc{\rho+\brc{1-\rho}\brc{\delta+\pi_1}}}, \\
  I_{23}=&\frac{m_{+1}\rho\brc{2\pi_1-1}}{\brc{1-\brc{1-\rho}\pi_1}\brc{\rho+\brc{1-\rho}\pi_1}}
  +\frac{m_{+2}\rho\brc{2\brc{\delta+\pi_1}-1}}{\brc{1-\brc{1-\rho}\brc{\delta+\pi_1}}\brc{\rho+\brc{1+\rho}\brc{\delta+\pi_1}}}, \\
  I_{31}=&I_{13}, \\
  I_{32}=&I_{23}, \\
  I_{33}=&\frac{m_{+2}\brc{1+\rho}\pi_1\brc{1-\pi_1}}{\brc{1-\rho}\brc{\rho+\brc{1-\rho}^2\pi_1\brc{1-\pi_1}}}
  +\frac{m_{+2}\brc{1+\rho}\brc{\delta+\pi_1}\brc{1-\delta-\pi_1}}{\brc{1-\rho}\brc{1-\brc{1-\rho}\brc{\delta+\pi_1}}\brc{\rho+\brc{1-\rho}\brc{\delta+\pi_1}}}.
\end{align*}

\vspace{2em}
\declare{Author contributions}{The authors confirm contribution to the paper as follows: study conception and design: Zhou J, Ma C-X; analysis and interpretation of results: Zhou J, Ma C-X; draft manuscript preparation: Zhou J. All authors reviewed the results and approved the final version of the manuscript.}

\declare{Conflict of interest}{The authors declare that they have no conflict of interest.}

\declare{Data availability}{The data presented in this study are openly available in references~\cite{mandel1982duration,liang2024homogeneity}.}

\bibliographystyle{unsrt}
\bibliography{correlateddata}

\end{document}